\documentclass[letter]{aa}
\usepackage{natbib} \usepackage{graphicx,color}
\bibpunct{(}{)}{;}{a}{}{,}
\begin{document}

\title{Transient horizontal magnetic fields in solar plage regions}

\author{R. Ishikawa\inst{1,2} \and S. Tsuneta\inst{2} \and K. Ichimoto\inst{2}
            \and H. Isobe\inst{3} \and Y. Katsukawa\inst{2} \and B. W. Lites\inst{4}
            \and S. Nagata\inst{5} \and T. Shimizu\inst{6} \and R. A. Shine\inst{7}
            \and Y. Suematsu\inst{2} \and T. D. Tarbell\inst{7} \and A. M. Title\inst{7}}

\institute{Department of Astronomy, University of Tokyo, Hongo, Bunkyo-ku, Tokyo 113-0033, Japan
 \and National Astronomical Observatory of Japan, 2-21-1 Osawa, Mitaka, Tokyo 181-8588, Japan
 \and Department of Earth and Planetary Science, University of Tokyo, Hongo, Bunkyo-ku, Tokyo, 113-0033, Japan
 \and High Altitude Observatory, National Center for Atmospheric Research, P.O. Box 3000, Boulder CO 80307-3000, USA
 \and Kwasan and Hida Observatories, Kyoto University, Yamashina, Kyoto, 607-8471, Japan
 \and ISAS/JAXA, Sagamihara, Kanagawa, 229-8510, Japan
 \and Lockheed Martin Solar and Astrophysics Laboratory, B/252, 3251 Hanover St., Palo Alto, CA 94304, USA}
\date{Received/Accepted}

\abstract
{}
{We report the discovery of isolated, small-scale emerging magnetic fields in a plage region with the Solar Optical Telescope aboard Hinode.}
{Spectro-polarimetric observations were carried out with a cadence of 34 seconds for the plage region located near disc center. The vector magnetic fields are inferred by Milne-Eddington inversion. }
{The observations reveal widespread occurrence of transient, spatially isolated horizontal magnetic fields. The lateral extent of the horizontal magnetic fields is comparable to the size of photospheric granules.
These horizontal magnetic fields seem to be tossed about by upflows and downflows of the granular convection.
We also report an event that appears to be driven by the magnetic buoyancy instability.
 We refer to buoyancy-driven emergence as type1 and convection-driven emergence as type2.
Although both events have magnetic field strengths of about 600 G, the filling factor of type1 is a factor of two larger than that of type2.}
{Our finding suggests that the granular convection in the plage regions is characterized by a high rate of occurrence of granular-sized transient horizontal fields.}

\maketitle
\section{Introduction}
Sunspots and active regions are believed to form from magnetic flux globally emerging from the solar interior into the solar atmosphere.
The emergence of magnetic flux is due to the buoyancy of large-scale, coherent toroidal flux tubes in the convection zone \citep{parker1955}.
Emerging horizontal fields become more or less vertical to the photosphere, and the vertical fields provide natural connectivity with observed X-ray loops in the corona.
This established paradigm is widely supported by observations and theory, but a concern has been raised on theoretical grounds that flux rising buoyantly through the convection zone will fail to maintain a coherent magnetic structure, and therefore fail to emerge \citep{magara2001,moreno1995}.
Here we report on observations at granular scales of widespread emergence of isolated magnetic fields that may be related to the concern.

Our observations were made with the Solar Optical Telescope \citep[SOT,][]{Tsuneta2007} aboard Hinode \citep{Kosugi2007}.
This telescope performs high-resolution ($\sim$0\farcs3) polarimetric observations with a very stable point spread function.
Its Spectro-Polarimeter (SP) obtains the complete polarization state (Stokes $I$, $Q$, $U$, and $V$) of the absorption profiles of the Zeeman-sensitive photospheric lines FeI~630.15~nm and FeI~630.25~nm.
We observed a plage region located near the center of the solar disk.
A narrow area of 2\farcs5 (8 slit positions, East-West) $\times$ 164\arcsec (North-South) was rapidly scanned with a coarse pixel size of 0\farcs32 by the SP, resulting in the moderate spatial resolution of $\sim$0\farcs6, and very high time cadence of 34 seconds over 40 minutes.
\begin{figure*}
\begin{center}
\includegraphics[width=6.5cm, angle=90]{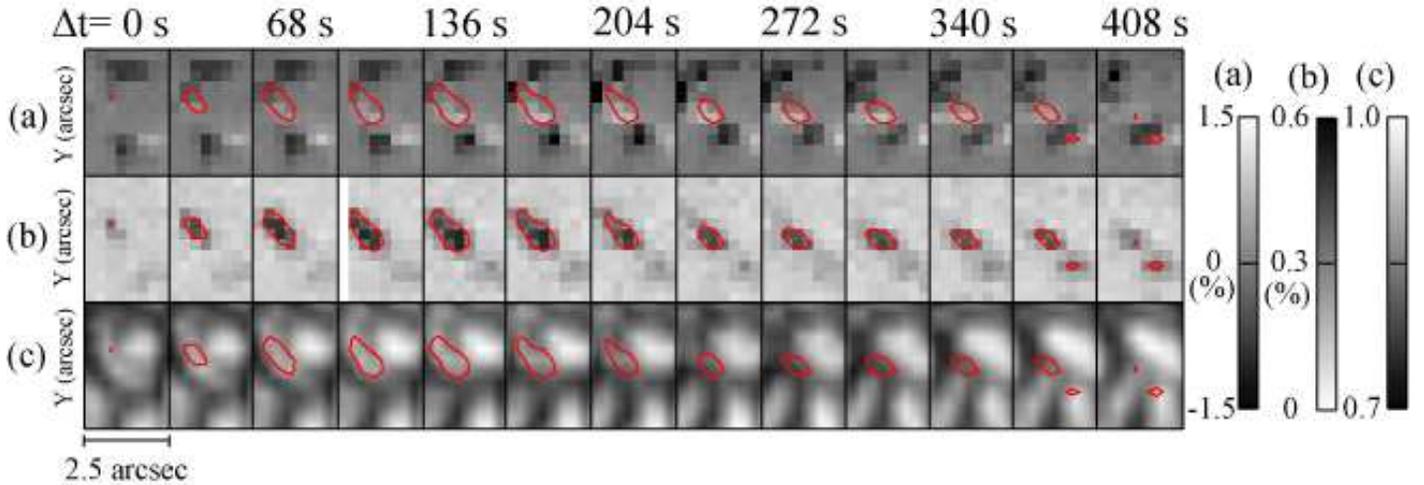}
\end{center}
\caption{The plage region located near disk center (72\arcsec W, 76\arcsec S) was observed from 16:00 to 16:40 UT on February 10, 2007. The Stokes profiles at two slit positions with integration time of 1.6 s each and scan step of 0\farcs16 each are summed, and 2 pixels along the slit with a pixel size of 0\farcs16 are also summed to obtain polarization accuracy better than 0.1\%: The images consist of 8 slit positions (0\farcs32 width) with a total scan time of 34 s. The pixel size along the slit is 0\farcs32. The evolution of physical quantities for the plage region are shown:
\textbf{(a)} $CP$ (vertical magnetic field),
\textbf{(b)} $LP$ (horizontal magnetic field),
\textbf{(c)} $I_{c}$.
The region where $LP$ is larger than 0.3\% is enclosed by red lines. The emergence of the horizontal magnetic flux starts at $\Delta t = 0$ s. Solar north is up and east to the left in all images of this report.
}
\label{type2_map}
\end{figure*}
  
\section{Ubiquitous transient fields in a plage region}
We obtain the continuum intensity, $I_{c}$, linear polarization, $LP=\int \sqrt{Q^{2}+U^{2}}d\lambda/\int I_{c} d\lambda$, and circular polarization, $CP=\int V d\lambda/\int I_{c} d\lambda$.
The integration for $LP$ is performed between -21.6 pm and +21.6 pm from the average center of the Stokes $I$ profiles (FeI~630.25~nm) for pixels without magnetic signals, and the integration for $CP$ is between +4.32 pm and +21.6 pm from the center of each Stokes $I$ profile. In the weak field limit, $LP$ is proportional to squares of the transverse field strength, and $CP$ to the line-of-sight field strength.

We discovered the frequent appearance and disappearance of nearly horizontal magnetic structures in the plage region: we find 52 occurrences of transient horizontal magnetic fields in this short time series.
These events appear to be randomly located spatially and temporally in the 2\farcs5 $\times$ 164\arcsec region observed for a duration of 40 min.
These events are not associated with existing long-lived magnetic fields, and appear in a region of insignificant vertical fields.
An episode of such transient horizontal magnetic fields is shown in Figure \ref{type2_map}.
The maximum size of the transient horizontal magnetic field is $\sim$1\farcs4 $\times$1\arcsec, and its duration is about 6 min.
A remarkable feature is that the horizontal magnetic structure first appears inside the granule, subsequently moves to the inter-granular lane, and finally disappears in the inter-granular lane.
The horizontal magnetic structure continues to be smaller than or at most as large as the size of the granule where it appears.
Horizontal magnetic fields with negative vertical magnetic fields at one end appear at $\Delta$t=34 s (black patch in Fig.\ref{type2_map}), where there are no pre-existing vertical magnetic fields@at $\Delta$t=0 s.
Vertical magnetic fields with positive polarity start to be seen clearly at the opposite end at $\Delta$t=102 s. 
After $\Delta$t=238 s, it is difficult to identify the bipolar configuration around the shrinking horizontal magnetic fields due to the weak vertical magnetic signals.

The transient horizontal fields discovered in the plage regions have lifetimes ranging from a minute to about ten minutes, comparable to the lifetime of granules.
Among 52 events, 43 horizontal magnetic structures appear inside the granules, and four appear in inter-granular lanes with the remaining five events ambiguous in position.
Since 52 events are detected in the 2\farcs5 $\times$ 164\arcsec observing area during
the 40 minutes, a new horizontal field appears every
46 seconds in the same observing region. The turnover time of the granules
is approximately 1000 sec with a velocity of 2 km s$^{-1}$ and with a depth
comparable to the horizontal scale.
There are approximately 182 granules in the observing area,
assuming that the size of the granules is 1\farcs5 $\times$ 1\farcs5.
84\% of these granules are not associated with stable strong vertical magnetic fields,
and we use this smaller sample for estimating the frequency of events.
Were every granule to have one embedded horizontal magnetic field
structure, the horizontal field would have appeared at the surface every 6.6
seconds ($\sim$1000 seconds/152 granules) in the observing area.
This shows that more than approximately 10\% of the granules have embedded horizontal fields, suggesting a relatively common occurrence of horizontal magnetic fields underneath the plage region.

We examined the Stokes profiles along the strong horizontal magnetic structure at $\Delta$t=136 s (Figure \ref{0210_profile}).
The Stokes $Q$ and $U$ profiles representing the horizontal magnetic fields are blue-shifted by 0-1 km s$^{-1}$ with respect to the line center of each Stokes $I$ profiles. The strong horizontal magnetic field rises along with the upward convective motion of the granule in the early phase of the event; these magnetic fields would be subject to the convective flows.
The polarity of the very weak vertical field component detected by the Stokes $V$ profile reverses along the horizontal magnetic structure, indicating either convex- or concave-upward magnetic fields, depending on the resolution of the 180-degree ambiguity in the direction of the horizontal field.
Because we follow the emergence of this event from its inception, the only logical configuration is convex-upward; i.e., an emerging omega-shaped loop.
This scenario is consistent with the higher upward flow near the center of the granular cell.

The transient horizontal fields reported here differ from the typical emerging flux driven by the Parker instability \citep[][and references therein]{parker1955,shibata1989,Strous1996,lites1998,Kubo2003}.
This difference can be seen in an observation made with SOT which we describe next.

\section{Comparison with buoyancy-driven emerging flux}
\begin{figure}
\begin{center}
\includegraphics[width=6cm, angle=90]{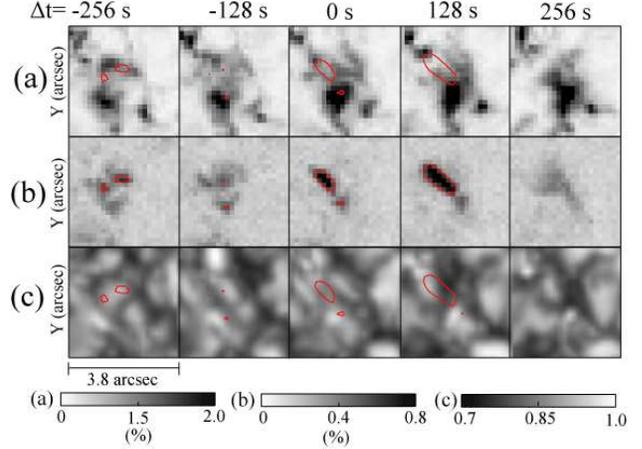}
\end{center}
\caption{NOAA AR 0931 located not far from disk center (289\arcsec E, 83\arcsec S) was observed from 17:35 to 18:25 UT on December 28, 2006. The images consist of 24 consecutive slit positions (0\farcs16 width) with a total scan time of 128 s. The pixel size along the slit is 0\farcs16.
\textbf{(a)} $nCP$ (vertical magnetic field),
\textbf{(b)} $LP$ (horizontal magnetic field),
\textbf{(c)} $I_{c}$.
The region where $LP$ is larger than 0.5\% is enclosed by red lines.}
\label{type1_map}
\end{figure}

\
We also present an example of an emerging flux event in a remnant active region located around the disk center. The properties of the event are considerably different from the previous example.
We find only one such event in the observing area of 3\farcs8 $\times$ $82\arcsec$ during 50 minutes' observation.
The episode starts with the appearance of horizontal magnetic fields, that appear to cause a considerable deformation of the surrounding granules (Figure \ref{type1_map}). The maximum size of the horizontal structure is $\sim$1\farcs6 along its major axis. At $\Delta$t=-128 s, prior to the emergence of magnetic flux, the granules become elongated with their alignment parallel to the orientation of the emerging fields. At $\Delta$t=0 s, strong horizontal magnetic field structure appears in the region between the elongated granules and existing vertical fields, and vertical magnetic components are detected at both ends of the horizontal magnetic structure.
Note that we employ unsigned circular polarization, $nCP=\int |V| d\lambda/\int I_{c} d\lambda$, instead of $CP$ because highly deformed Stokes $V$ profiles, some of which have three lobes, are seen in the region. The integration is carried out with the same wavelength range as $LP$.
The horizontal flux emerges with separating vertical components at both ends. At $\Delta$t=128 s, the horizontal magnetic structure becomes longer and stronger, and the vertical magnetic fields also strengthen.
The bright point appears where the vertical field is strong. In the next frame ($\Delta$t= 256 s), the horizontal magnetic structure disappears (most likely rising above the line formation height), and we are left with vertical bipolar components.
The series of observations indicate that the emerging magnetic field structure is stiff against the granular motion since it affects the surrounding granules even before its emergence into the photosphere.
      
The Stokes profiles along the horizontal magnetic structure at $\Delta$t=128 s are shown in Figure \ref{1228_profile}. The Stokes $Q$ and $U$ profiles show that the direction of the magnetic vector is roughly parallel to the major axis of the flux tube.
The Stokes $U$ profiles representing the horizontal magnetic structure are blue-shifted by 1-2 km s$^{-1}$ at the top of the loop with respect to the line center of the Stokes $I$ profiles, while the Stokes $V$ profile is strongly red-shifted at one end of the horizontal magnetic structure with a velocity up to 5 km s$^{-1}$, which is below but close to the local sound speed (6-7 km s$^{-1}$).
The strong downward flow may be related to the convective collapse \citep{Bellot2001}. The expansion velocity is estimated to be about 2 km s$^{-1}$ from the change of the length of the horizontal magnetic structure. This velocity is comparable to the upward velocity at the top of the loop.
The high-velocity downflows at one end of the flux tube continuing over 4 minutes show
that plasma descends in the flux tube in association with the rising flux tube.
At the other footpoint, a few km s$^{-1}$ upflow is detected mainly at $\Delta$t=128 s.
This may be a reverse flow due to the local enhancement of gas pressure triggered by a downflow near the footpoint.
Another possibility is siphon flow driven by the pressure gradient of the two foot points \citep{CargillandPriest1980}.
Note that even with the upflow at one end of the footpoints, the higher downflow dominates, and reduces the net mass in the flux tube. The magnetic buoyancy will be sustained as a result. This property is closer to that of emerging flux by the Parker instability than the previous one.

\begin{figure}
\begin{center}
\includegraphics[width=5.5cm, angle=90]{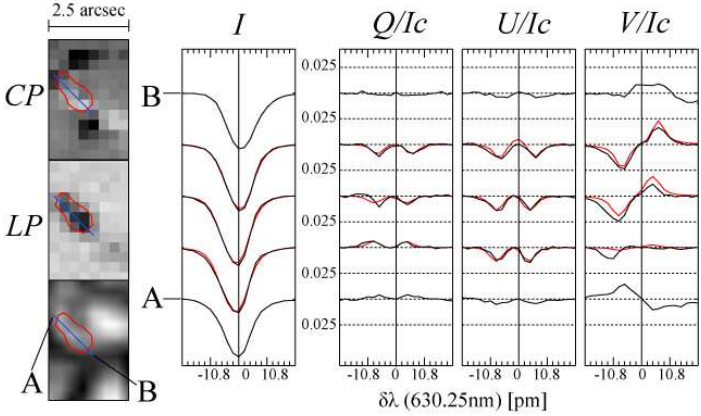}
\end{center}
\caption{The frame at $\Delta t=$ 136 sec in Fig.\ref{type2_map} is enlarged on the left. Stokes profiles ($I, Q, U$, and $V$) along the major axis (blue line) are shown to the right with black lines. The fitted profiles are shown with red lines. The profiles marked by A and B correspond to the NE and SW ends of the horizontal field structure, respectively. The vertical line in the Stokes profiles shows the average center of the Stokes $I$ absorption profiles (FeI~630.25~nm) without magnetic fields, and provides a fiducial for the zero point in velocity. The tick mark indicates 2.16 pm.}
\label{0210_profile}
\end{figure}

\begin{figure}
\begin{center}
\includegraphics[width=5.5cm, angle=90]{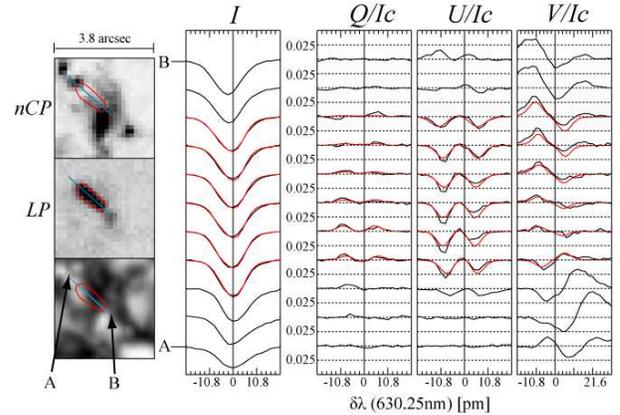}
\end{center}
\caption{The frame at $\Delta t=$ 128 sec in Fig.\ref{type1_map} enlarged on the left.
The caption is the same as Fig.\ref{0210_profile}.
}
\label{1228_profile}
\end{figure}

Given the significant difference in the two episodes presented here, we hereafter call this episode type1, and the previous episode type2. We do not find any events of type1 in the data set where we find many type2 events. The occurrence rate of type1 is much lower than that of type2.
The three components of the magnetic field are estimated by a Milne-Eddington inversion for the Stokes spectra shown in the Figures \ref{0210_profile} and \ref{1228_profile}. (The black lines indicate the observed Stokes spectra, while the red lines the fitted spectra.) The average horizontal magnetic field strengths are 560 Gauss (type1) and 580 Gauss (type2). The field lines are roughly aligned along the major axis of the strong horizontal magnetic field region.
For type2, we have chosen the event having the strongest $LP$, so that we were able to perform a reliable Stokes inversion. Other type2 events generally have smaller $LP$. 

A remarkable difference between the types 1 and 2 is in the average filling factors, which are 0.44 and 0.17, respectively.
The filling factor is an areal fraction of the magnetic atmosphere within a resolution element.
We estimate the maximum diameter of the flux tube under the assumption of a single horizontal flux tube partially occupying a pixel.
The width is equal to the size of the telescope's point spread function (0\farcs32) multiplied by the filling factor.
The size of the type1 horizontal flux tube thus derived is less than $\sim$100 km, while that for the type2 is less than $\sim$40 km.
They are much smaller than the scale height of the line forming layer, indicating that actual diameter of the flux tube is not as small as this.
Nevertheless, it is true that the type2 tube is much thinner than the type1 tube.

\section{Discussion}
The occurrence of isolated small-scale, transient horizontal magnetic fields in the quiet Sun was discovered using spectro-polarimetric data from the Advanced Stokes Polarimeter at a spatial resolution of 1\arcsec \citep{Lites1996}. These horizontal fields are interpreted to be convection-driven or buoyancy-driven emerging bipolar fields. Very recently, the high time variability and ubiquity of horizontal fields in the quiet Sun were clearly shown in data from the SOLIS and GONG instruments \citep{Harvey2007}. Subsequently, Hinode data has permitted us to directly see the ubiquitous presence of horizontal magnetic fields in the quiet Sun at unprecedented small scales \citep{Lites2008,Centeno2007,orozco2007}.
It is as yet unclear if the granular-size emergence phenomenon in the plage region reported herein pertains to flux elements having a stronger degree of polarization than much of that reported for the quiet Sun. It is important to understand whether the plage and quiet Sun horizontal fields have a similar origin.

The equi-partition field strength $B_{e}$ is the field strength where magnetic energy is equal to kinetic energy:
$ \frac{B_{e}^{2}}{8\pi}=\frac{1}{2}\rho\upsilon^{2}$.
The typical equipartition field strength $B_{e}$ is about 400 Gauss for granules with a velocity of $\upsilon = 2\times10^{5}$ cm s$^{-1}$, and the density $\rho$ is $3\times10^{-7}$ g cm$^{-2}$ at the solar surface. The magnetic field strengths of both type1 and type2 are apparently stronger than the equi-partition field.
The magnetic field strength of type1 higher than the equi-partition
field strength is consistent with the observed robustness against 
granular motion \citep{Cheung2007}. The type1 event is interpreted as essentially the instability-driven
emergence of flux tubes.
As strong a field for type2 cannot be readily explained.
It appears that the horizontal flux tubes of the type 2 event are carried by the upflow of convection to the photosphere, and then the horizontal flow of the convection transports them to the inter-granular lane. The type2 event is apparently subject to granular motion.

The buoyant force is proportional to $(\rho_{e}-\rho_{i}) \pi g a^{2}$, where $\rho_{e}$ and $\rho_{i}$ are the plasma density inside and outside the horizontal flux tubes, and $a$ the radius of the flux tube. The drag force is given by $0.5 C_{d} \rho_{e}u^{2} a$, where $C_{d}$ is an aerodynamic coefficient of order unity, and $u$ the turbulent velocity. The magnetic tension force is $\pi a^{2} B^{2}/4\pi L$, where $L$ is the curvature radius of horizontal fields.
We substitute a granular size of 10$^{3}$ km for $L$ to obtain the upper limit of magnetic tension.
The three forces become comparable for $a$ of 50 km and $B$ of 600 G.
The type2 tube diameter of 40 km means that the effect of buoyancy is relatively smaller than the drag force.
This implies that a thinner flux tube is more strongly influenced by turbulent motion.

Convective collapse \citep{Parker1978} would not work due to its horizontal nature throughout the evolution for the type 2 event.
Indeed, \citet{Abbett2007} shows that a convective surface dynamo can generate horizontally directed magnetic structures.
The transient horizontal fields revealed by Hinode in our study suggest the possibility of such a local dynamo process \citep{Cattaneo1999,Vogler2007}.
An alternative physical picture is presented by the extended fields caused by an exploding magnetic structure \citep{moreno1995} and/or by horizontal magnetic fields that failed to emerge \citep{magara2001}.

\begin{acknowledgements}
The authors thank B. C. Low for his valuable comments and encouragement.
Hinode is a Japanese mission developed and launched by ISAS/JAXA, with NAOJ as a domestic partner and NASA and STFC (UK) as international partners. It is operated by these agencies in co-operation with ESA and NSC (Norway). The National Center for Atmospheric Research is sponsored by the National Science Foundation.\end{acknowledgements}
\bibliographystyle{aa}
\bibliography{9022}
\end{document}